\documentclass[%
 reprint,
 amsmath,amssymb,
 aps,
 superscriptaddress,
 prl,
 twocolumn
]{revtex4}

\usepackage{graphicx}
\usepackage{dcolumn}
\usepackage{bm}

\usepackage{xcolor}

\begin{document}

\preprint{APS/123-QED}

\title{Engineered swift equilibration for arbitrary geometries}

\author{Adam G. Frim}%
\author{Adrianne Zhong}%
\author{Shi-Fan Chen}
\author{Dibyendu Mandal}
\affiliation{%
 Department of Physics, University of California, Berkeley, Berkeley, CA, 94720
}%
\author{Michael R. DeWeese}
\affiliation{%
 Department of Physics, University of California, Berkeley, Berkeley, CA, 94720
}%
\affiliation{%
Redwood Center For Theoretical Neuroscience and Helen Wills Neuroscience Institute, University of California, Berkeley, Berkeley, CA, 94720
}%

\date{\today}

\begin{abstract}
Engineered swift equilibration (ESE) is a class of driving protocols that enforce an equilibrium distribution with respect to external control parameters at the beginning and end of rapid state transformations of open, classical non-equilibrium systems. ESE protocols have previously been derived and experimentally realized for Brownian particles in simple, one-dimensional, time-varying trapping potentials; one recent study considered ESE in two-dimensional Euclidean configuration space. Here we extend the ESE framework to generic, overdamped Brownian systems in arbitrary curved configuration space and illustrate our results with specific examples not amenable to previous techniques. Our approach may be used to impose the necessary dynamics to control the full temporal configurational distribution in a wide variety of experimentally realizable settings.

\end{abstract}

\maketitle


\indent \textit{Introduction.}---In any transformation process, there exists some intrinsic relaxation time for the final distribution to be reached. Recently, a number of studies have attempted to manipulate or eliminate altogether this relaxation time by means of alternative driving protocols. These strategies are generally known as shortcuts to adiabaticity, in which one attempts to rapidly transform from a specified initial distribution to a target distribution at a specified final time, in either classical~\cite{2013_PRA_Jarzynski,2014_PRX_Deffner,2017_NJP_Patra,2019_RMP_Guery-Odelin} or quantum~\cite{2000_PRL_Emmanouilidou,2009_JPhysA_Berry,2013_Torrontegui,2013_PRL_del_Campo,2013_PRA_Jarzynski,2014_PRX_Deffner,2016_NC_an,2017_NJP_Patra,Funo_PRL_2017,2018_Lutz,2019_RMP_Guery-Odelin,2020_Lutz} settings. In the context of open classical systems, protocols that shortcut the natural relaxation timescale of the system go by the name of Engineered Swift Equilibration (ESE) and focus on enforcing internal thermal equilibrium at specified initial and final times~\cite{2016_Nature_ESE,2018_PRE_Chupeau}. This constraint is clearly satisfied when an instantaneous equilibrium distribution is maintained at \textit{all times} during the protocol, rather than only at the beginning and the end, a strategy called shortcuts to isothermality introduced in~\cite{2017_PRE_Li} on which we focus. This is achieved by adding external driving forces: by following a specified driving protocol, a rapidly-transforming system assumes the trajectory of a quasistatic transformation. 

To be more concrete, consider a physical system described by some time-dependent Hamiltonian $H_0(\lambda_i(t))$, where $t$ is time, and all time dependence is prescribed by parameters $\lambda_i(t)$. Following standard Boltzmann statistics~\cite{Kadanoff}, the equilibrium probability distribution at a given time is 
\begin{equation} \label{eq:equilibrium}
    \rho_\mathrm{eq}(\mathbf{x}; \lambda_i(t)) = \frac{\exp(-\beta H_0(\mathbf{x}; \lambda_i(t)))}{Z(\lambda_i(t))},
\end{equation}
where $\beta = (k_BT)^{-1}$ is the inverse temperature, $k_B$ is Boltzmann constant, and $Z(\lambda_i(t))$ is the partition function, explicitly dependent on parameters $\lambda_i(t)$. If the $\{\lambda_i(t)\}$ are changed quasistatically, the system will be well-described by Boltzmann statistics at all times. However, if $\lambda_i(t)$ changes sufficiently rapidly, the system deviates from its equilibrium distribution specified by Eq.~(\ref{eq:equilibrium}). The ESE protocols we consider introduce a modified Hamiltonian $H(\mathbf{x},t) = H_0(\mathbf{x},\lambda_i(t))+H_1(\mathbf{x},t)$ such that under the full dynamics of $H$, the system assumes the internal equilibrium distribution of $H_0$ alone [Eq.~(\ref{eq:equilibrium})], at all times.

To date, ESE protocols have been successfully derived for a Brownian particle trapped in a variety of simple one-dimensional (1D) potentials~\cite{2017_PRE_Li, 2018_NJP_Chupeau} and realized experimentally for a Brownian particle in a 1D harmonic trap~\cite{2016_Nature_ESE}. This approach was also recently applied to the Brownian gyrator~\cite{2020_PRE_Baldassarri}, a two-dimensional (Euclidean) system in contact with two heat baths that admits non-equilibrium steady states. In this letter, we extend the ESE framework to generic overdamped Brownian systems, including those with arbitrarily high dimensional, non-Euclidean configuration spaces. We demonstrate the utility of our framework by numerically finding the ESE forcing for previously unsolved systems. Due to the wide applicability of overdamped Brownian dynamics, we expect our results to prove useful to the many physical contexts where swift, controlled transitions are often highly desired, such as nanoscale engineering \cite{2007_PRL_Schmiedl,2007_EPL_Schmiedl}, thermodynamic computing \cite{2014_erasure_PRE_Zulkowski, 2018_arxiv_boyd}, and manipulating colloidal systems \cite{2017_SM_Martinez, 2011_NatPhys_Blickle}, to name just a few.

\indent \textit{Theory.} We consider a particle undergoing Brownian motion in the overdamped limit whose dynamics are governed by the Langevin equation,
\begin{equation}
\label{eq:Langevin}
    \gamma \frac{d\mathbf{x}(t)}{dt} = - \mathbf{\nabla} V(\mathbf{x}(t); \lambda_i(t)) 
    + \bm{\eta}(t) + \mathbf{F}_{\mathrm{ext}}(t),
\end{equation} %
\noindent
where $\mathbf{x}$ is the position of the particle, $\gamma$ is the viscosity, $V(\mathbf{x}; \lambda_i(t))$ is the potential acting on the particle parameterized by control parameters $\lambda_i(t)$, $\bm{\eta}$ is Gaussian noise with delta function autocorrelation $\langle \eta_i(t) \eta_j(t') \rangle = (2\gamma/\beta)\delta_{ij}\delta(t - t')$, where $i,j$ index Euclidean coordinates and $\mathbf{F}_{\mathrm{ext}}(t)$ is an external force on the particle. 

Following standard procedures~\cite{Kadanoff}, this leads to a Fokker-Planck equation:
\begin{equation} \label{eq:FP}
    \partial_t \rho(\mathbf{x}, t) = \mathbf{\nabla} \cdot \bigg[ \bigg(\frac{\mathbf{\nabla} V(\mathbf{x}; \lambda_i(t)) - \mathbf{F}_\mathrm{ext}}{\gamma} + \frac{\mathbf{\nabla} }{\beta \gamma} \bigg) \rho(\mathbf{x}, t) \bigg],
\end{equation}
where $\rho(\mathbf{x}, t)$ is the configuration space probability distribution at a given time. In the absence of an external force, the steady state solution is found by setting the LHS of Eq.~\eqref{eq:FP} to zero, yielding the usual Boltzmann distribution, $\rho_\mathrm{eq}$.

Now suppose that the control parameters $\lambda_i(t)$ are time-dependent and varied too quickly to assume a quasistatic transition. We seek $\mathbf{F}_\mathrm{ext}(t)$ such that  $\rho(\mathbf{x}(t), t; \lambda_i(t)) =\rho_\text{eq}(\mathbf{x}; \lambda_i(t))$ for all times $ t$. This can only be satisfied if all explicitly time-dependent terms in Eq.~(\ref{eq:FP}) independently cancel. Defining $\mathbf{P} \equiv \rho_\mathrm{eq}\mathbf{F}_\mathrm{ext}$, this constraint can be written
\begin{equation} \label{eq:ESE}
\nabla \cdot \mathbf{P} = -\gamma \partial_t \rho_\mathrm{eq} .
\end{equation}%
We now generalize to systems whose configuration space is an arbitrary compact Riemannian manifold $M$ with metric $g$, and write the vector $\mathbf{P}$ as a differential 1-form $P = P_i \mathrm{d} x^{i}$. Equation~(\ref{eq:ESE}) then generalizes to
\begin{equation} \label{eq:ESEdiff}
  \mathrm{d}^{\dagger} P = -\gamma \partial_t \rho_\text{eq},
\end{equation}
where $\mathrm{d}$ is the exterior derivative and $\mathrm{d}^\dagger$ is its Hodge dual, whose action on a $k$-form is given: $\mathrm{d}^\dagger = (-1)^{n(k-1)+1}\star \mathrm{d}\star$, where $\star$ is the Hodge star operator and $n$ is the dimension of the manifold. We now invoke the Hodge Decomposition, which states that, for any $k$-form $P_k$ on $M$, there exists a unique decomposition~\cite{Nakahara}:
\begin{equation} \label{eq:hodgeD}
    P_k = \mathrm{d} A_{k-1} + \mathrm{d}^{\dagger}B_{k+1} + C_k ,
\end{equation}
where $A_{k-1}$ and $B_{k+1}$ are $(k-1)$- and $(k+1)$-forms, respectively, and $C_k$ is a harmonic $k$-form; {\em i.e.}  $\Delta C_k = 0$, and $\Delta = \mathrm{d}\mathrm{d}^{\dagger} + \mathrm{d}^{\dagger} \mathrm{d}$ is the generalized Laplace operator. On a compact Riemannian manifold, harmonic $k$-forms also satisfy $\mathrm{d}C_k = \mathrm{d}^\dagger C_k = 0$ \cite{voisin_2002}. Therefore, we may write Eq.~(\ref{eq:ESE}) as
\begin{equation} \label{eq:ESEHodge}
  \mathrm{d}^{\dagger} P = \mathrm{d}^{\dagger}(\mathrm{d} A + \mathrm{d}^{\dagger}B + C)= \mathrm{d}^{\dagger} \mathrm{d} A + \mathrm{d}^{\dagger}\mathrm{d}^{\dagger}B + \mathrm{d}^{\dagger} C = \Delta A,
\end{equation}
where $A$ is a 0-form, $B$ is a 2-form, and $C$ is a harmonic 1-form. Eq.~\eqref{eq:ESEdiff} thus becomes a generalized Poisson's equation:
\begin{equation} \label{eq:ESEManifold}
  \Delta A = -\gamma \partial_t \rho_{\mathrm{eq}} .
\end{equation}
Coupling this with Eq.~(\ref{eq:hodgeD}) and taking $B=C=0$ for simplicity yields our main result:
\begin{equation} \label{eq:ESEManifoldForce}
  \mathbf{F}_{\text{ext}} = \frac{\mathrm{d} A}{\rho_{\mathrm{eq}}}.
\end{equation}
We now demonstrate the utility of our result by applying it to multiple physically realizable examples.

\indent \textit{Euclidean Configuraton Space.}---In $d$-dimensional Euclidean space, the Hodge decomposition trivializes to the Helmholtz decomposition~\cite{Nakahara}. Our results still hold, though for certain cases the decomposition may no longer be unique. For this space, the generalized Laplace operator is simply the standard Laplace operator. Note that our choice $B = C = 0$ amounts to a no-curl gauge: $\nabla \times \mathbf{P} = 0$. We may thus define a scalar potential $A$ that satisfies $\nabla A = \mathbf{P}$. This is analogous to standard electrostatics, where $A$ and $\mathbf{P}$ play the roles of the electric potential and field, respectively~\cite{Jackson}.

Given the general solution to the Laplace operator for Euclidean space, we have, for $d\neq 2$,
\begin{equation}
A(\mathbf{x}) = -\frac{\Gamma(\frac{d-2}{2})}{4\pi^\frac{d}{2}}\int d^d\mathbf{x}'  \ |\mathbf{x}-\mathbf{x}'|^{2-d}(-\gamma \partial_t\rho_{\text{eq}}(\mathbf{x}'))
\end{equation}
and $\mathbf{F}_{\text{ext}}(\mathbf{x}) =\rho^{-1}_{\text{eq}} \mathbf{\nabla} A$. Note that this solution reproduces Eq.~(12) of ~\cite{2017_PRE_Li} for 1D systems. For the special case of $d=2$, the potential is given by
\begin{equation}
A(\mathbf{x}) = \frac{1}{2\pi}\int d^2\mathbf{x}'  \ \log |\mathbf{x}-\mathbf{x}'|(-\gamma \partial_t\rho_{\text{eq}}(\mathbf{x}')).
\end{equation}

\begin{figure*}[t]
\label{fig:ESEfig}
    \centering
    \includegraphics[scale=0.9]{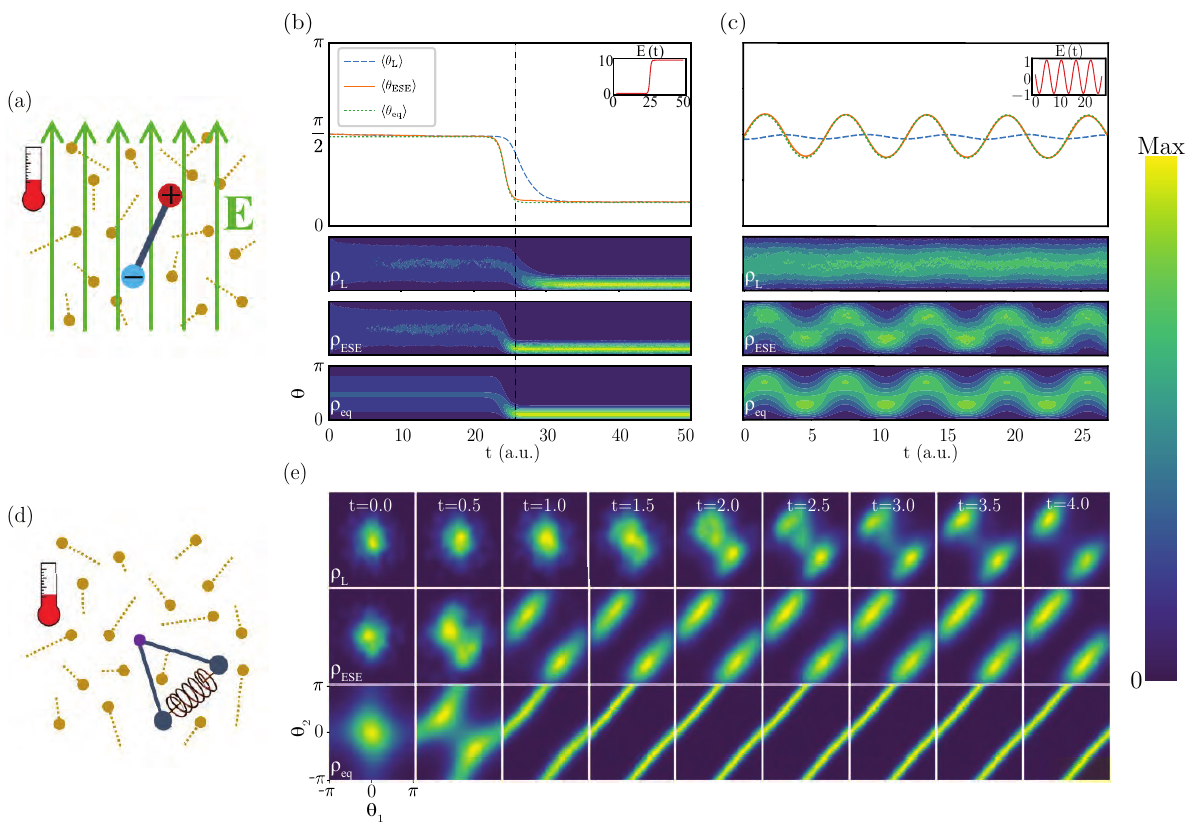}
    \caption{Simulations demonstrate that ESE protocols produce densities closely tracking the equilibrium distribution in configuration space corresponding to the control parameters at each moment in time. \textbf{(a)} Schematic diagram of the ensemble considered in subplots \textbf{(b)} and \textbf{(c)}. An ensemble of electric dipoles are placed in a uniform, time-varying electric field. Due to these constraints, the state of a dipole is specified by a polar coordinate $\theta$ and an azimuthal angle $\phi$ such that the system's configuration space is a (2D) sphere. \textbf{(b)} The electric field is varied sigmoidally in time, as shown in the inset. In the top panel, we show the average values of $\theta$ over the entire ensemble for a system undergoing Langevin dynamics (dashed blue), ESE dynamics (solid orange), and for a Boltzmann distibution for the given electric field $E(t)$, denoted $\rho_{\text{eq}}$ (dotted green). The three lower panels are the corresponding full distributions for $\theta$ at all times. \textbf{(c)} Same as panel \text{b} for the sinusoidally varying electric field shown in the inset. \textbf{(d)} Schematic diagram of the ensemble considered in subplot \textbf{(e)}. An ensemble of two coupled pendula with time-varying coupling constant. The state of a coupled pendulum is specified by the two angular coordinates of the pendula, $\theta_1$, $\theta_2 \in [-\pi,\pi)$, such that the full system's configuration space is a torus. \textbf{(e)} The probability distributions of an ensemble of coupled pairs of pendula undergoing Langevin dynamics (top) and ESE dynamics (middle) plotted against the instantaneous Boltzmann distribution (bottom) for a coupling constant that changes from zero at small times to a negative value.
    }
\end{figure*}

\indent \textit{Spherical Configuration Space.}---We now consider topologically nontrivial configuration spaces. Our first example is an electric dipole with dipole moment $\mathbf{p} = p\hat{\mathbf{p}}$ placed in a time-varying electric field pointed in the $z$-direction, $\mathbf{E} = E(t) \hat{\mathbf{z}}$, as shown in Fig.~1(a). The potential energy for this system is 
\begin{equation}
\label{eq:dipole}
    V(\theta, \phi, t)  = -p E(t) \cos (\theta),
\end{equation}
where $\cos\theta = \hat{\mathbf{p}}\cdot \hat{\mathbf{z}}$ and $\phi$ measures the azimuthal angle about the 
$z$-axis.
The configuration space of this system is the 2-sphere, $M = S^2$, for which the Laplace operator is
\begin{equation}
    \Delta_{S^2}  = \frac{1}{\sin \theta} \frac{\partial}{\partial \theta} \bigg(\sin \theta \frac{\partial }{\partial \theta} \bigg) + \frac{\partial^2 }{\partial \phi^2}.
\end{equation}
To calculate the required ESE force, we find $\rho_{\text{eq}}(t)$, the instantaneous Boltzmann distribution for this system: 
\begin{equation}
\rho_{\text{eq}}(t) = \left[\frac{4\pi \sinh(\beta pE(t))}{\beta p E(t)}\right]^{-1} \exp{\beta pE(t)\cos \theta},
\end{equation}
such that the governing equation is given by $\Delta_{S^2} A = -\gamma \partial_t \rho_\text{eq}$. In the high temperature limit, one may find an explicit expression for $A$ (see Supplemental Materials Section SM.1); however, finding a closed-form expression is, in general, intractable. Instead, we employ a series expansion in the spherical harmonics, $Y_{\ell}^m(\theta,\phi)$. The spherical harmonics are the eigenfunctions of the spherical Laplace operator, {\em i.e.} $\Delta_{S^2}Y_{\ell}^m(\theta,\phi) = -\ell(\ell+1)Y_{\ell}^m(\theta,\phi) $, such that if we write $\rho_{\text{eq}} = \sum_{\ell,m}c_{\ell,m}(t)Y_\ell^m(\theta,\phi)$, then by the orthogonality and completeness of the spherical harmonics, we have
\begin{equation}
\label{eq:SphericalHarms}
A = \gamma\sum_{\ell=0}^\infty \sum_{m=-\ell}^\ell\frac{\partial_t c_{\ell,m}(t)}{\ell(\ell+1)}Y_\ell^m(\theta,\phi).
\end{equation}
Note that $c_{\ell,m}$ may likewise be computed:
\begin{equation}
\label{eq:SphericalHarmsCoeffs}
c_{\ell,m}(t) = \int_\Omega \rho_{\text{eq}}(\theta,\phi,t) Y_\ell^{m*}(\theta,\phi)d\Omega.
\end{equation}
Due to the azimuthal symmetry of $\rho_{\text{eq}}(\theta,\phi,t)$, only $m=0$ terms will be nonzero, simplifying our analysis. Finally, 
$\mathbf{F}_\text{ext}$ is found by taking $\mathbf{P} = \mathrm{d} A \implies \mathbf{F}_\text{ext} = \rho_\text{eq}^{-1} (\mathrm{\nabla}_{S^2}) A$.

We now simulate the system for specified functions $E(t)$. The dynamics of this system are governed by a set of Langevin equations:
\begin{align}
&m(\ddot{\theta}-\sin\theta\cos\theta \dot{\phi}^2)=-\gamma \dot{\theta} -pE\cos\theta+\eta_\theta+F_{\text{ext},\theta} , \label{eq:sphericaldynamics1}\\
&m(\sin\theta \ddot{\phi}+2\cos\theta\dot{\theta}\dot{\phi})=-\gamma \sin\theta \dot{\phi}+\eta_\phi+F_{\text{ext},\phi}. \label{eq:sphericaldynamics2}
\end{align}
Note that, though we do not explicitly enforce the overdamped limit in Eqs.~(\ref{eq:sphericaldynamics1}) and (\ref{eq:sphericaldynamics2}), we will effectively do so by means of parameter choices in our simulations. For a specified $E(t)$, we numerically solve Eqs.~(\ref{eq:SphericalHarms}) and (\ref{eq:SphericalHarmsCoeffs}) to find the ESE force (truncating above $\ell=5$) and then simulate the Langevin dynamics in both the presence and absence of this force for an ``ensemble" of $10^4$ dipoles. In units of $\beta = m = p = 1$, we simulate with $\gamma = 20$. Due to the noise terms in Eqs.~(\ref{eq:sphericaldynamics1}) and (\ref{eq:sphericaldynamics2}), these are stochastic differential equations, which we simulate by means of a first-order Euler-Maruyama algorithm~\cite{Kloeden} with step size of $dt = 0.01$ time units. Given the promotion of configuration space to a non-Euclidean manifold, the relation for the noise term is modified~\cite{2009_EPL_Kumar,2017_PRE_apaza}: $\langle \eta_i(t) \eta_j(t') \rangle =(2\gamma/\beta)g^{ij}\delta(t-t') $, where $g^{ij}$ is the inverse metric of the manifold. For the (unit) sphere, the inverse metric is $g^{\theta\theta} = 1$, $g^{\phi\phi} = 1/\sin\theta$, and all other entries are zero. Therefore, following the standard Euler-Maruyama treatment, we take $\eta_\theta dt = \sqrt{2\gamma dt/\beta}\mathcal{N}(0,1)$ and $\eta_\phi dt = \sqrt{2\gamma dt \csc\theta/\beta}\mathcal{N}(0,1)$, where $\mathcal{N}(0,1)$ is the Normal distribution with zero mean and unit variance. To deal with the spherical-polar coordinate singularities at $\theta = 0$ and $\theta = \pi$, we temporarily rotate to a different local coordinate system and numerically integrate a single time-step whenever $0<\theta<\pi/10$ or $9\pi/10<\theta<\pi$ (see~\cite{SM}).

In Fig.~1 we plot both the mean value and the full probability distribution of $\theta(t)$ for both the standard Langevin and ESE dynamics for two representative, temporally-varying electric fields, as described below. Due to the azimuthal symmetry of the problem, the distribution over the azimuthal angle $\phi$ is not affected by any temporal change in $E(t)$.

In Fig.~1(b), we consider a sigmoidally varying electric field, $E(t) \sim E_0 + (\Delta E) S(t - t_c)$, where 
\begin{equation} \label{eq:logistic}
S(t) = (1 + e^{-t})^{-1}
\end{equation}
is the logistic function, $\Delta E = 10$ is the amplitude of the change of the electric field, and $t_c$ is the transition time. For $E_0$ small, 
$\theta$ is primarily distributed about $\pi/2$, which corresponds to the equator. We note that for $pE_0 \ll k_B T$, the dipoles have no preferred direction, so they should be uniformly distributed throughout configuration space. However, the distribution appears non-uniform as a function of $\theta$. This is an artifact of our coordinate system: there is more phase space area at $\theta = \pi/2$ (the equator) than elsewhere, such that the probability as a function of $\theta$ should be non-uniform. For $t\gg t_c$, we find that, on average, $\theta<\pi/2$. This is physically sensible: the electric field is strong and directed along the $z$-axis such that the dipoles will tend to align with it. However, another artifact of our coordinate system is the absence of probability density at the pole at $\theta = 0$. For $t\sim t_c$, in the absence of the ESE force, the system remains out of equilibrium for a finite period of time before eventually relaxing to the new equilibrium. However, when the ESE force is introduced, the system remains close to the equilibrium distribution at all times.

In Fig 1(c), we consider a sinusoidal electric field, $E(t)\sim E\sin(t)$. In this case, we see that the constantly changing field never allows the standard Langevin system to fully equilibrate; instead, the system oscillates with an approximate $\pi/2$ phase-shift at a significantly smaller amplitude. Conversely, the ESE dynamics converge close to the equilibrium distribution at all points in time. For this case, we point out one subtlety: due to the periodicity of the drive, there exists a non-equilibrium periodic steady-state distribution over $\theta$ in the absence of ESE forcing, to which the Langevin dynamics converge \cite{Pankratov_1, Pankratov_2}. However, reaching this steady-state is not the goal of our ESE protocol. Rather, we seek to track the instantaneous Boltzmann distribution for the temporally varying external control parameters at all times, even in this periodic case.

\indent \textit{Toroidal configuration space.}---Next, we consider toroidal configuration space, as exemplified by a system of two pendula, each of mass $m$ and unit length suspended vertically, and coupled to each other with a time-varying coupling constant $\kappa(t)$, as illustrated in Fig.~1(d). The potential may be modeled as
 \begin{equation}
    V(\theta_1, \theta_2, t) \simeq -(mg\cos\theta_1+mg\cos\theta_2-\kappa(t) \cos(\theta_1-\theta_2)),
\end{equation}
where $\theta_1$ and $\theta_2$ are angles of the respective pendula with respect to the $z$-axis. Given the periodicity in $\theta_i$, the configuration space of this system is the 2-torus, $M = T^2 = [0,2\pi]\times [0,2\pi]$. The Laplace operator for this manifold is
\begin{equation}
    \Delta_{T^2}  =  \frac{\partial^2}{\partial\theta_1^2}+\frac{\partial^2}{\partial\theta_2^2} , 
   \end{equation}
where one must recall the periodicity of the coordinates: $\theta_i \sim \theta_i+2\pi n$ for $n\in \mathbb{Z}$. We again employ a series expansion to solve Eq.~(\ref{eq:ESE}). In this case, we carry out a 2D Fourier series. Considering that $\Delta_{T^2} \exp{i(m_1\theta_1 + m_2\theta_2)} = -(m_1^2+m_2^2)\exp{i(m_1\theta_1 + m_2\theta_2)}$, we may write $\rho_\text{eq} = \sum_{m_1,m_2} c_{m_1,m_2}(t) \exp{i(m_1\theta_1 + m_2\theta_2)}$ and deduce that
\begin{equation}
	A = \gamma \sum_{m_1,m_2} \frac{\partial_tc_{m_1,m_2}(t)}{m_1^2+m_2^2} e^{i(m_1\theta_1 + m_2\theta_2)}.
 \end{equation}
We again simulate this dynamical system for a given $\kappa(t)$. The governing Langevin equations are now 
\begin{multline}
m\ddot{\theta}_{1/2}= -\gamma \dot{\theta}_{1/2}-mg\sin\theta_{1/2} \\ + \kappa(t)\sin(\theta_{1/2}-\theta_{2/1})+\eta_{\theta_{1/2}}+F_{\text{ext},\theta_{1/2}} . \label{eq:Toroiddynamics}\\
\end{multline}
As with the last example, the dynamics are not confined to the overdamped limit. For a specified $\kappa(t)$, we numerically solve the ESE force 
$F_{\text{ext}}(t)$ by means of the Fourier series expansion (truncating above $m_1=m_2=10$) and then simulate the Langevin dynamics in both the presence and absence of this force. In units of $\beta = m = g = 1$, we again simulate for $\gamma = 20$ and employ an Euler-Maruyama algorithm with time step $dt = 0.01$. Conveniently, the noise terms for this system do not have any geometric corrections, such that $\eta_{\theta_1}dt = \eta_{\theta_2} dt = \sqrt{2\gamma dt/\beta}\mathcal{N}(0,1)$. For our simulations, we choose $\kappa(t) \sim -\kappa_0 S(t)$, where $S(t)$ is defined by (\ref{eq:logistic}) so that the pendula are initially uncoupled but after some critical time $t_c$ they are anti-coupled. In Fig.~1(e), we display the resulting probability distribution for the coordinates $(\theta_1,\theta_2)$ at several times near $t_c$. In the absence of the ESE force (top row), equilibration happens over a finite amount of time as the system relaxes to its new anti-coupled distribution. However, when the ESE force is added (second row), the resulting distribution agrees well with the calculated equilibrium distribution $\rho_\text{eq}(t)$ corresponding to the control parameter values at each moment in time (bottom row).\\
\indent
\indent \textit{Discussion.}---In previous studies, the notions of optimality and control often refer to specific protocols designed to minimize excess work or some other performance index when changing between two equilibria or non-equilibrium steady states in finite time~\cite{2007_PRL_Schmiedl, 2007_EPL_Schmiedl,2012_PRL_Sivak,2012_PRE_zulkowski,2013_plos_zulkowski_optimal,2015_PRE_zulkowski,2019_PRE_Plata}. The ESE framework  may also be considered a control strategy, though ESE seeks only to minimize time to equilibration throughout the protocol without any constraints or penalties on the work required to do so. Prior work has analyzed the relation between the duration $\tau$ of a protocol and the energy dissipated in carrying out a drive, concluding that for a variety of model systems, the energy dissipated is proportional to $1/\tau$~\cite{esposito2010finite,Esposito_2013,2014_erasure_PRE_Zulkowski,2017_PRE_Li,2017_PRL_Campbell}. We may conjecture that the type of ESE protocols we have derived here will behave similarly, with dissipated energy scaling as $\tau_\text{relax}/\tau$, where $\tau_\text{relax}$ is the intrinsic viscous relaxation timescale for the overdamped system in consideration.\\
\indent
In principle, provided a smooth trajectory of control parameters, ESE should allow for arbitrarily rapid equilibrium switching of ensemble distributions. However, the difficulty of realizing the required forces in a laboratory setting would presumably preclude such a situation. In addition, the theory itself breaks down in such a limit due to higher order effects ignored in a basic Langevin treatment, such as a finite characteristic timescale of the noise correlations. Nonetheless, for the range of timescales for which Eq.~(\ref{eq:Langevin}) applies, ESE yields a method to achieve controlled, swift equilibration.\\
\indent
Our ESE protocol ensures a high degree of control throughout the drive. Not only do we enforce the mean, or the mean and variance (or any finite combination of moments) of the probability distribution, we dictate the \textit{entire} probability distribution at all times during the protocol. In fact, following~\cite{Nakahara}, if a scalar field integrates to zero over a full compact manifold, it may be written as the divergence of a vector field. Importantly, by integrating the LHS of Eq.~(\ref{eq:ESE}) over any phase space manifold $M$, we see that
\begin{equation}
    \int_M (-\gamma \partial_t \rho_\text{eq}) dV = -\gamma \frac{d}{dt} \int_M \rho_\text{eq} dV = -\gamma \frac{d}{dt} (1) = 0
\end{equation}
following conservation of probability. We conclude that for any arbitrary time-dependent potential described by some smooth set of coordinates, there will always be a corresponding ESE force that can enforce swift equilibration.\\
\indent
Finally, we note further degrees of freedom in the ESE condition defined by Eq.~(\ref{eq:ESEHodge}): The differential form $d^\dagger P = \Delta A$ allows for an alternative, arbitrary choice of a harmonic 1-form $C$ and a 2-form $B$ by using Eq.~(\ref{eq:hodgeD}). For a given trajectory specified by $\rho_{\text{eq}}(t)$, these choices lead to a class of inequivalent, though perhaps non-conservative \cite{2018_arxiv_boyd}, driving forces---where now we must use the full form of Eq.~\eqref{eq:hodgeD} $\rho_\text{eq} F =  \mathrm{d}A+\mathrm{d}^\dagger B +C$---each of which enforces swift equilibration. These additional degrees of freedom, which are inaccessible in low-dimensional Euclidean spaces and therefore have not been observed in past studies, afford greater flexibility in constructing appropriate forces for practical laboratory applications.\\
\indent
\indent \textit{Conclusion.}---In this letter, we have successfully extended the ESE protocol to systems with nontrivial configuration space topology. We hope our results will be useful for designing optimal strategies for manipulating a thermalized system of multiple canonical position variables swiftly through controlled parameter changes. Our methods can be used to calculate the necessary auxiliary forces to impose internal equilibrium dynamics in experimental settings, though the derivation we present here is only valid for the overdamped limit. In future work, it will be interesting to generalize our framework to include underdamped systems.

\begin{acknowledgments} AGF is supported by the NSF GRFP under Grant No. DGE 1752814. SC is supported by NSF GRFP under Grant No. DGE 1106400 and the Berkeley Astrophysics Center Astronomy and Astrophysics Graduate Fellowship.  MRD and DM were supported in part by the U. S. Army Research Laboratory and the U. S. Army Research Office under contracts W911NF-13-1-0390 (MRD and DM) and W911NF-20-1-0151 (MRD).
\end{acknowledgments}

\bibliography{ms}

\begin{thebibliography}{41}
\expandafter\ifx\csname natexlab\endcsname\relax\def\natexlab#1{#1}\fi
\expandafter\ifx\csname bibnamefont\endcsname\relax
  \def\bibnamefont#1{#1}\fi
\expandafter\ifx\csname bibfnamefont\endcsname\relax
  \def\bibfnamefont#1{#1}\fi
\expandafter\ifx\csname citenamefont\endcsname\relax
  \def\citenamefont#1{#1}\fi
\expandafter\ifx\csname url\endcsname\relax
  \def\url#1{\texttt{#1}}\fi
\expandafter\ifx\csname urlprefix\endcsname\relax\def\urlprefix{URL }\fi
\providecommand{\bibinfo}[2]{#2}
\providecommand{\eprint}[2][]{\url{#2}}

\bibitem[{\citenamefont{Jarzynski}(2013)}]{2013_PRA_Jarzynski}
\bibinfo{author}{\bibfnamefont{C.}~\bibnamefont{Jarzynski}},
  \bibinfo{journal}{Phys. Rev. A} \textbf{\bibinfo{volume}{88}},
  \bibinfo{pages}{040101(R)} (\bibinfo{year}{2013}),
  \urlprefix\url{https://link.aps.org/doi/10.1103/PhysRevA.88.040101}.

\bibitem[{\citenamefont{Deffner et~al.}(2014)\citenamefont{Deffner, Jarzynski,
  and del Campo}}]{2014_PRX_Deffner}
\bibinfo{author}{\bibfnamefont{S.}~\bibnamefont{Deffner}},
  \bibinfo{author}{\bibfnamefont{C.}~\bibnamefont{Jarzynski}},
  \bibnamefont{and} \bibinfo{author}{\bibfnamefont{A.}~\bibnamefont{del
  Campo}}, \bibinfo{journal}{Phys. Rev. X} \textbf{\bibinfo{volume}{4}},
  \bibinfo{pages}{021013} (\bibinfo{year}{2014}),
  \urlprefix\url{https://link.aps.org/doi/10.1103/PhysRevX.4.021013}.

\bibitem[{\citenamefont{Patra and Jarzynski}(2017)}]{2017_NJP_Patra}
\bibinfo{author}{\bibfnamefont{A.}~\bibnamefont{Patra}} \bibnamefont{and}
  \bibinfo{author}{\bibfnamefont{C.}~\bibnamefont{Jarzynski}},
  \bibinfo{journal}{New Journal of Physics} \textbf{\bibinfo{volume}{19}},
  \bibinfo{pages}{125009} (\bibinfo{year}{2017}),
  \urlprefix\url{https://doi.org/10.1088%2F1367-2630%2Faa924c}.

\bibitem[{\citenamefont{Gu\'ery-Odelin
  et~al.}(2019)\citenamefont{Gu\'ery-Odelin, Ruschhaupt, Kiely, Torrontegui,
  Mart\'{\i}nez-Garaot, and Muga}}]{2019_RMP_Guery-Odelin}
\bibinfo{author}{\bibfnamefont{D.}~\bibnamefont{Gu\'ery-Odelin}},
  \bibinfo{author}{\bibfnamefont{A.}~\bibnamefont{Ruschhaupt}},
  \bibinfo{author}{\bibfnamefont{A.}~\bibnamefont{Kiely}},
  \bibinfo{author}{\bibfnamefont{E.}~\bibnamefont{Torrontegui}},
  \bibinfo{author}{\bibfnamefont{S.}~\bibnamefont{Mart\'{\i}nez-Garaot}},
  \bibnamefont{and} \bibinfo{author}{\bibfnamefont{J.~G.} \bibnamefont{Muga}},
  \bibinfo{journal}{Rev. Mod. Phys.} \textbf{\bibinfo{volume}{91}},
  \bibinfo{pages}{045001} (\bibinfo{year}{2019}),
  \urlprefix\url{https://link.aps.org/doi/10.1103/RevModPhys.91.045001}.

\bibitem[{\citenamefont{Emmanouilidou et~al.}(2000)\citenamefont{Emmanouilidou,
  Zhao, Ao, and Niu}}]{2000_PRL_Emmanouilidou}
\bibinfo{author}{\bibfnamefont{A.}~\bibnamefont{Emmanouilidou}},
  \bibinfo{author}{\bibfnamefont{X.-G.} \bibnamefont{Zhao}},
  \bibinfo{author}{\bibfnamefont{P.}~\bibnamefont{Ao}}, \bibnamefont{and}
  \bibinfo{author}{\bibfnamefont{Q.}~\bibnamefont{Niu}},
  \bibinfo{journal}{Phys. Rev. Lett.} \textbf{\bibinfo{volume}{85}},
  \bibinfo{pages}{1626} (\bibinfo{year}{2000}),
  \urlprefix\url{https://link.aps.org/doi/10.1103/PhysRevLett.85.1626}.

\bibitem[{\citenamefont{Berry}(2009)}]{2009_JPhysA_Berry}
\bibinfo{author}{\bibfnamefont{M.~V.} \bibnamefont{Berry}},
  \bibinfo{journal}{Journal of Physics A: Mathematical and Theoretical}
  \textbf{\bibinfo{volume}{42}}, \bibinfo{pages}{365303}
  (\bibinfo{year}{2009}),
  \urlprefix\url{https://doi.org/10.1088%2F1751-8113%2F42%2F36%2F365303}.

\bibitem[{\citenamefont{Torrontegui et~al.}(2013)\citenamefont{Torrontegui,
  Ib{\'a}{\~n}ez, Mart{\'\i}nez-Garaot, Modugno, del Campo, Gu{\'e}ry-Odelin,
  Ruschhaupt, Chen, and Muga}}]{2013_Torrontegui}
\bibinfo{author}{\bibfnamefont{E.}~\bibnamefont{Torrontegui}},
  \bibinfo{author}{\bibfnamefont{S.}~\bibnamefont{Ib{\'a}{\~n}ez}},
  \bibinfo{author}{\bibfnamefont{S.}~\bibnamefont{Mart{\'\i}nez-Garaot}},
  \bibinfo{author}{\bibfnamefont{M.}~\bibnamefont{Modugno}},
  \bibinfo{author}{\bibfnamefont{A.}~\bibnamefont{del Campo}},
  \bibinfo{author}{\bibfnamefont{D.}~\bibnamefont{Gu{\'e}ry-Odelin}},
  \bibinfo{author}{\bibfnamefont{A.}~\bibnamefont{Ruschhaupt}},
  \bibinfo{author}{\bibfnamefont{X.}~\bibnamefont{Chen}}, \bibnamefont{and}
  \bibinfo{author}{\bibfnamefont{J.~G.} \bibnamefont{Muga}}, in
  \emph{\bibinfo{booktitle}{Advances in Atomic, Molecular, and Optical
  Physics}}, edited by
  \bibinfo{editor}{\bibfnamefont{E.}~\bibnamefont{Arimondo}},
  \bibinfo{editor}{\bibfnamefont{P.~R.} \bibnamefont{Berman}},
  \bibnamefont{and} \bibinfo{editor}{\bibfnamefont{C.~C.} \bibnamefont{Lin}}
  (\bibinfo{publisher}{Academic Press}, \bibinfo{year}{2013}),
  vol.~\bibinfo{volume}{62}, pp. \bibinfo{pages}{117 -- 169},
  \urlprefix\url{http://www.sciencedirect.com/science/article/pii/B9780124080904000025}.

\bibitem[{\citenamefont{del Campo}(2013)}]{2013_PRL_del_Campo}
\bibinfo{author}{\bibfnamefont{A.}~\bibnamefont{del Campo}},
  \bibinfo{journal}{Phys. Rev. Lett.} \textbf{\bibinfo{volume}{111}},
  \bibinfo{pages}{100502} (\bibinfo{year}{2013}),
  \urlprefix\url{https://link.aps.org/doi/10.1103/PhysRevLett.111.100502}.

\bibitem[{\citenamefont{An et~al.}(2016)\citenamefont{An, Lv, del Campo, and
  Kim}}]{2016_NC_an}
\bibinfo{author}{\bibfnamefont{S.}~\bibnamefont{An}},
  \bibinfo{author}{\bibfnamefont{D.}~\bibnamefont{Lv}},
  \bibinfo{author}{\bibfnamefont{A.}~\bibnamefont{del Campo}},
  \bibnamefont{and} \bibinfo{author}{\bibfnamefont{K.}~\bibnamefont{Kim}},
  \bibinfo{journal}{Nature Communications} \textbf{\bibinfo{volume}{7}},
  \bibinfo{pages}{12999} (\bibinfo{year}{2016}),
  \urlprefix\url{https://doi.org/10.1038/ncomms12999}.

\bibitem[{\citenamefont{Funo et~al.}(2017)\citenamefont{Funo, Zhang, Chatou,
  Kim, Ueda, and del Campo}}]{Funo_PRL_2017}
\bibinfo{author}{\bibfnamefont{K.}~\bibnamefont{Funo}},
  \bibinfo{author}{\bibfnamefont{J.-N.} \bibnamefont{Zhang}},
  \bibinfo{author}{\bibfnamefont{C.}~\bibnamefont{Chatou}},
  \bibinfo{author}{\bibfnamefont{K.}~\bibnamefont{Kim}},
  \bibinfo{author}{\bibfnamefont{M.}~\bibnamefont{Ueda}}, \bibnamefont{and}
  \bibinfo{author}{\bibfnamefont{A.}~\bibnamefont{del Campo}},
  \bibinfo{journal}{Phys. Rev. Lett.} \textbf{\bibinfo{volume}{118}},
  \bibinfo{pages}{100602} (\bibinfo{year}{2017}),
  \urlprefix\url{https://link.aps.org/doi/10.1103/PhysRevLett.118.100602}.

\bibitem[{\citenamefont{Abah and Lutz}(2018)}]{2018_Lutz}
\bibinfo{author}{\bibfnamefont{O.}~\bibnamefont{Abah}} \bibnamefont{and}
  \bibinfo{author}{\bibfnamefont{E.}~\bibnamefont{Lutz}},
  \bibinfo{journal}{Phys. Rev. E} \textbf{\bibinfo{volume}{98}},
  \bibinfo{pages}{032121} (\bibinfo{year}{2018}),
  \urlprefix\url{https://link.aps.org/doi/10.1103/PhysRevE.98.032121}.

\bibitem[{\citenamefont{Abah et~al.}(2020)\citenamefont{Abah, Paternostro, and
  Lutz}}]{2020_Lutz}
\bibinfo{author}{\bibfnamefont{O.}~\bibnamefont{Abah}},
  \bibinfo{author}{\bibfnamefont{M.}~\bibnamefont{Paternostro}},
  \bibnamefont{and} \bibinfo{author}{\bibfnamefont{E.}~\bibnamefont{Lutz}},
  \bibinfo{journal}{Phys. Rev. Research} \textbf{\bibinfo{volume}{2}},
  \bibinfo{pages}{023120} (\bibinfo{year}{2020}),
  \urlprefix\url{https://link.aps.org/doi/10.1103/PhysRevResearch.2.023120}.

\bibitem[{\citenamefont{Mart{\'\i}nez et~al.}(2016)\citenamefont{Mart{\'\i}nez,
  Petrosyan, Gu{\'e}ry-Odelin, Trizac, and Ciliberto}}]{2016_Nature_ESE}
\bibinfo{author}{\bibfnamefont{I.~A.} \bibnamefont{Mart{\'\i}nez}},
  \bibinfo{author}{\bibfnamefont{A.}~\bibnamefont{Petrosyan}},
  \bibinfo{author}{\bibfnamefont{D.}~\bibnamefont{Gu{\'e}ry-Odelin}},
  \bibinfo{author}{\bibfnamefont{E.}~\bibnamefont{Trizac}}, \bibnamefont{and}
  \bibinfo{author}{\bibfnamefont{S.}~\bibnamefont{Ciliberto}},
  \bibinfo{journal}{Nature Physics} \textbf{\bibinfo{volume}{12}},
  \bibinfo{pages}{843} (\bibinfo{year}{2016}),
  \urlprefix\url{https://doi.org/10.1038/nphys3758}.

\bibitem[{\citenamefont{Chupeau
  et~al.}(2018{\natexlab{a}})\citenamefont{Chupeau, Besga, Gu\'ery-Odelin,
  Trizac, Petrosyan, and Ciliberto}}]{2018_PRE_Chupeau}
\bibinfo{author}{\bibfnamefont{M.}~\bibnamefont{Chupeau}},
  \bibinfo{author}{\bibfnamefont{B.}~\bibnamefont{Besga}},
  \bibinfo{author}{\bibfnamefont{D.}~\bibnamefont{Gu\'ery-Odelin}},
  \bibinfo{author}{\bibfnamefont{E.}~\bibnamefont{Trizac}},
  \bibinfo{author}{\bibfnamefont{A.}~\bibnamefont{Petrosyan}},
  \bibnamefont{and}
  \bibinfo{author}{\bibfnamefont{S.}~\bibnamefont{Ciliberto}},
  \bibinfo{journal}{Phys. Rev. E} \textbf{\bibinfo{volume}{98}},
  \bibinfo{pages}{010104(R)} (\bibinfo{year}{2018}{\natexlab{a}}),
  \urlprefix\url{https://link.aps.org/doi/10.1103/PhysRevE.98.010104}.

\bibitem[{\citenamefont{Li et~al.}(2017)\citenamefont{Li, Quan, and
  Tu}}]{2017_PRE_Li}
\bibinfo{author}{\bibfnamefont{G.}~\bibnamefont{Li}},
  \bibinfo{author}{\bibfnamefont{H.~T.} \bibnamefont{Quan}}, \bibnamefont{and}
  \bibinfo{author}{\bibfnamefont{Z.~C.} \bibnamefont{Tu}},
  \bibinfo{journal}{Phys. Rev. E} \textbf{\bibinfo{volume}{96}},
  \bibinfo{pages}{012144} (\bibinfo{year}{2017}),
  \urlprefix\url{https://link.aps.org/doi/10.1103/PhysRevE.96.012144}.

\bibitem[{\citenamefont{Kadanoff}(2000)}]{Kadanoff}
\bibinfo{author}{\bibfnamefont{L.}~\bibnamefont{Kadanoff}},
  \emph{\bibinfo{title}{Statics, Dynamics and Renormalization}}
  (\bibinfo{publisher}{World Scientific Publishing Company},
  \bibinfo{year}{2000}).

\bibitem[{\citenamefont{Chupeau
  et~al.}(2018{\natexlab{b}})\citenamefont{Chupeau, Ciliberto,
  Gu{\'{e}}ry-Odelin, and Trizac}}]{2018_NJP_Chupeau}
\bibinfo{author}{\bibfnamefont{M.}~\bibnamefont{Chupeau}},
  \bibinfo{author}{\bibfnamefont{S.}~\bibnamefont{Ciliberto}},
  \bibinfo{author}{\bibfnamefont{D.}~\bibnamefont{Gu{\'{e}}ry-Odelin}},
  \bibnamefont{and} \bibinfo{author}{\bibfnamefont{E.}~\bibnamefont{Trizac}},
  \bibinfo{journal}{New Journal of Physics} \textbf{\bibinfo{volume}{20}},
  \bibinfo{pages}{075003} (\bibinfo{year}{2018}{\natexlab{b}}),
  \urlprefix\url{https://doi.org/10.1088%2F1367-2630%2Faac875}.

\bibitem[{\citenamefont{Baldassarri et~al.}(2020)\citenamefont{Baldassarri,
  Puglisi, and Sesta}}]{2020_PRE_Baldassarri}
\bibinfo{author}{\bibfnamefont{A.}~\bibnamefont{Baldassarri}},
  \bibinfo{author}{\bibfnamefont{A.}~\bibnamefont{Puglisi}}, \bibnamefont{and}
  \bibinfo{author}{\bibfnamefont{L.}~\bibnamefont{Sesta}},
  \bibinfo{journal}{Phys. Rev. E} \textbf{\bibinfo{volume}{102}},
  \bibinfo{pages}{030105(R)} (\bibinfo{year}{2020}),
  \urlprefix\url{https://link.aps.org/doi/10.1103/PhysRevE.102.030105}.

\bibitem[{\citenamefont{Schmiedl and
  Seifert}(2007{\natexlab{a}})}]{2007_PRL_Schmiedl}
\bibinfo{author}{\bibfnamefont{T.}~\bibnamefont{Schmiedl}} \bibnamefont{and}
  \bibinfo{author}{\bibfnamefont{U.}~\bibnamefont{Seifert}},
  \bibinfo{journal}{Phys. Rev. Lett.} \textbf{\bibinfo{volume}{98}},
  \bibinfo{pages}{108301} (\bibinfo{year}{2007}{\natexlab{a}}),
  \urlprefix\url{https://link.aps.org/doi/10.1103/PhysRevLett.98.108301}.

\bibitem[{\citenamefont{Schmiedl and
  Seifert}(2007{\natexlab{b}})}]{2007_EPL_Schmiedl}
\bibinfo{author}{\bibfnamefont{T.}~\bibnamefont{Schmiedl}} \bibnamefont{and}
  \bibinfo{author}{\bibfnamefont{U.}~\bibnamefont{Seifert}},
  \bibinfo{journal}{{EPL} (Europhysics Letters)} \textbf{\bibinfo{volume}{81}},
  \bibinfo{pages}{20003} (\bibinfo{year}{2007}{\natexlab{b}}),
  \urlprefix\url{https://doi.org/10.1209%2F0295-5075%2F81%2F20003}.

\bibitem[{\citenamefont{Zulkowski and
  DeWeese}(2014)}]{2014_erasure_PRE_Zulkowski}
\bibinfo{author}{\bibfnamefont{P.~R.} \bibnamefont{Zulkowski}}
  \bibnamefont{and} \bibinfo{author}{\bibfnamefont{M.~R.}
  \bibnamefont{DeWeese}}, \bibinfo{journal}{Phys. Rev. E}
  \textbf{\bibinfo{volume}{89}}, \bibinfo{pages}{052140}
  (\bibinfo{year}{2014}),
  \urlprefix\url{https://link.aps.org/doi/10.1103/PhysRevE.89.052140}.

\bibitem[{\citenamefont{Boyd et~al.}(2018)\citenamefont{Boyd, Patra, Jarzynski,
  and Crutchfield}}]{2018_arxiv_boyd}
\bibinfo{author}{\bibfnamefont{A.~B.} \bibnamefont{Boyd}},
  \bibinfo{author}{\bibfnamefont{A.}~\bibnamefont{Patra}},
  \bibinfo{author}{\bibfnamefont{C.}~\bibnamefont{Jarzynski}},
  \bibnamefont{and} \bibinfo{author}{\bibfnamefont{J.~P.}
  \bibnamefont{Crutchfield}}, \emph{\bibinfo{title}{Shortcuts to thermodynamic
  computing: The cost of fast and faithful erasure}} (\bibinfo{year}{2018}),
  \eprint{arXiv: 1812.11241}.

\bibitem[{\citenamefont{Mart{\'\i}nez et~al.}(2017)\citenamefont{Mart{\'\i}nez,
  Rold{\'a}n, Dinis, and Rica}}]{2017_SM_Martinez}
\bibinfo{author}{\bibfnamefont{I.~A.} \bibnamefont{Mart{\'\i}nez}},
  \bibinfo{author}{\bibfnamefont{{\'E}.}~\bibnamefont{Rold{\'a}n}},
  \bibinfo{author}{\bibfnamefont{L.}~\bibnamefont{Dinis}}, \bibnamefont{and}
  \bibinfo{author}{\bibfnamefont{R.~A.} \bibnamefont{Rica}},
  \bibinfo{journal}{Soft Matter} \textbf{\bibinfo{volume}{13}},
  \bibinfo{pages}{22} (\bibinfo{year}{2017}),
  \urlprefix\url{http://dx.doi.org/10.1039/C6SM00923A}.

\bibitem[{\citenamefont{Blickle and Bechinger}(2012)}]{2011_NatPhys_Blickle}
\bibinfo{author}{\bibfnamefont{V.}~\bibnamefont{Blickle}} \bibnamefont{and}
  \bibinfo{author}{\bibfnamefont{C.}~\bibnamefont{Bechinger}},
  \bibinfo{journal}{Nature Physics} \textbf{\bibinfo{volume}{8}},
  \bibinfo{pages}{143} (\bibinfo{year}{2012}),
  \urlprefix\url{https://doi.org/10.1038/nphys2163}.

\bibitem[{\citenamefont{Nakahara}(2003)}]{Nakahara}
\bibinfo{author}{\bibfnamefont{M.}~\bibnamefont{Nakahara}},
  \emph{\bibinfo{title}{Geometry, Topology and Physics}}
  (\bibinfo{publisher}{CRC Press}, \bibinfo{year}{2003}),
  \bibinfo{edition}{2nd} ed.

\bibitem[{\citenamefont{Voisin}(2002)}]{voisin_2002}
\bibinfo{author}{\bibfnamefont{C.}~\bibnamefont{Voisin}},
  \emph{\bibinfo{title}{Harmonic Forms and Cohomology}}
  (\bibinfo{publisher}{Cambridge University Press}, \bibinfo{year}{2002}),
  vol.~\bibinfo{volume}{1} of \emph{\bibinfo{series}{Cambridge Studies in
  Advanced Mathematics}}, pp. \bibinfo{pages}{117--136}.

\bibitem[{\citenamefont{Jackson}(1998)}]{Jackson}
\bibinfo{author}{\bibfnamefont{J.~D.} \bibnamefont{Jackson}},
  \emph{\bibinfo{title}{Classical Electrodynamics, Third Edition}}
  (\bibinfo{publisher}{Wiley}, \bibinfo{year}{1998}), ISBN
  \bibinfo{isbn}{9780471309321}.

\bibitem[{\citenamefont{Kloeden and Platen}(1992)}]{Kloeden}
\bibinfo{author}{\bibfnamefont{P.~E.} \bibnamefont{Kloeden}} \bibnamefont{and}
  \bibinfo{author}{\bibfnamefont{E.}~\bibnamefont{Platen}},
  \emph{\bibinfo{title}{Numerical Solution of Stochastic Differential
  Equations}} (\bibinfo{publisher}{Springer Berlin Heidelberg},
  \bibinfo{year}{1992}),
  \urlprefix\url{https://doi.org/10.1007/978-3-662-12616-5}.

\bibitem[{\citenamefont{Kumar and Kumar}(2009)}]{2009_EPL_Kumar}
\bibinfo{author}{\bibfnamefont{N.}~\bibnamefont{Kumar}} \bibnamefont{and}
  \bibinfo{author}{\bibfnamefont{K.~V.} \bibnamefont{Kumar}},
  \bibinfo{journal}{{EPL} (Europhysics Letters)} \textbf{\bibinfo{volume}{86}},
  \bibinfo{pages}{17001} (\bibinfo{year}{2009}),
  \urlprefix\url{https://doi.org/10.1209%2F0295-5075%2F86%2F17001}.

\bibitem[{\citenamefont{Apaza and Sandoval}(2017)}]{2017_PRE_apaza}
\bibinfo{author}{\bibfnamefont{L.}~\bibnamefont{Apaza}} \bibnamefont{and}
  \bibinfo{author}{\bibfnamefont{M.}~\bibnamefont{Sandoval}},
  \bibinfo{journal}{Phys. Rev. E} \textbf{\bibinfo{volume}{96}},
  \bibinfo{pages}{022606} (\bibinfo{year}{2017}),
  \urlprefix\url{https://link.aps.org/doi/10.1103/PhysRevE.96.022606}.

\bibitem[{SM()}]{SM}
\bibinfo{note}{See Supplemental Material at [URL will be inserted by publisher]
  for details of a high temperature expansion, further simulation details, and
  a movie of a given simulation run as in Fig. 1(e).}

\bibitem[{\citenamefont{Pankratov and Salerno}(2000)}]{Pankratov_1}
\bibinfo{author}{\bibfnamefont{A.~L.} \bibnamefont{Pankratov}}
  \bibnamefont{and} \bibinfo{author}{\bibfnamefont{M.}~\bibnamefont{Salerno}},
  \bibinfo{journal}{Phys. Rev. E} \textbf{\bibinfo{volume}{61}},
  \bibinfo{pages}{1206} (\bibinfo{year}{2000}),
  \urlprefix\url{https://link.aps.org/doi/10.1103/PhysRevE.61.1206}.

\bibitem[{\citenamefont{Pankratov}(2002)}]{Pankratov_2}
\bibinfo{author}{\bibfnamefont{A.~L.} \bibnamefont{Pankratov}},
  \bibinfo{journal}{Phys. Rev. E} \textbf{\bibinfo{volume}{65}},
  \bibinfo{pages}{022101} (\bibinfo{year}{2002}),
  \urlprefix\url{https://link.aps.org/doi/10.1103/PhysRevE.65.022101}.

\bibitem[{\citenamefont{Sivak and Crooks}(2012)}]{2012_PRL_Sivak}
\bibinfo{author}{\bibfnamefont{D.~A.} \bibnamefont{Sivak}} \bibnamefont{and}
  \bibinfo{author}{\bibfnamefont{G.~E.} \bibnamefont{Crooks}},
  \bibinfo{journal}{Phys. Rev. Lett.} \textbf{\bibinfo{volume}{108}},
  \bibinfo{pages}{190602} (\bibinfo{year}{2012}),
  \urlprefix\url{https://link.aps.org/doi/10.1103/PhysRevLett.108.190602}.

\bibitem[{\citenamefont{Zulkowski et~al.}(2012)\citenamefont{Zulkowski, Sivak,
  Crooks, and DeWeese}}]{2012_PRE_zulkowski}
\bibinfo{author}{\bibfnamefont{P.~R.} \bibnamefont{Zulkowski}},
  \bibinfo{author}{\bibfnamefont{D.~A.} \bibnamefont{Sivak}},
  \bibinfo{author}{\bibfnamefont{G.~E.} \bibnamefont{Crooks}},
  \bibnamefont{and} \bibinfo{author}{\bibfnamefont{M.~R.}
  \bibnamefont{DeWeese}}, \bibinfo{journal}{Physical Review E}
  \textbf{\bibinfo{volume}{86}}, \bibinfo{pages}{041148}
  (\bibinfo{year}{2012}), ISSN \bibinfo{issn}{1539-3755, 1550-2376},
  \urlprefix\url{https://link.aps.org/doi/10.1103/PhysRevE.86.041148}.

\bibitem[{\citenamefont{Zulkowski et~al.}(2013)\citenamefont{Zulkowski, Sivak,
  and DeWeese}}]{2013_plos_zulkowski_optimal}
\bibinfo{author}{\bibfnamefont{P.~R.} \bibnamefont{Zulkowski}},
  \bibinfo{author}{\bibfnamefont{D.~A.} \bibnamefont{Sivak}}, \bibnamefont{and}
  \bibinfo{author}{\bibfnamefont{M.~R.} \bibnamefont{DeWeese}},
  \bibinfo{journal}{PLoS ONE} \textbf{\bibinfo{volume}{8}},
  \bibinfo{pages}{e82754} (\bibinfo{year}{2013}), ISSN
  \bibinfo{issn}{1932-6203},
  \urlprefix\url{https://dx.plos.org/10.1371/journal.pone.0082754}.

\bibitem[{\citenamefont{Zulkowski and DeWeese}(2015)}]{2015_PRE_zulkowski}
\bibinfo{author}{\bibfnamefont{P.~R.} \bibnamefont{Zulkowski}}
  \bibnamefont{and} \bibinfo{author}{\bibfnamefont{M.~R.}
  \bibnamefont{DeWeese}}, \bibinfo{journal}{Physical Review E}
  \textbf{\bibinfo{volume}{92}}, \bibinfo{pages}{032117}
  (\bibinfo{year}{2015}), ISSN \bibinfo{issn}{1539-3755, 1550-2376},
  \urlprefix\url{https://link.aps.org/doi/10.1103/PhysRevE.92.032117}.

\bibitem[{\citenamefont{Plata et~al.}(2019)\citenamefont{Plata, Gu\'ery-Odelin,
  Trizac, and Prados}}]{2019_PRE_Plata}
\bibinfo{author}{\bibfnamefont{C.~A.} \bibnamefont{Plata}},
  \bibinfo{author}{\bibfnamefont{D.}~\bibnamefont{Gu\'ery-Odelin}},
  \bibinfo{author}{\bibfnamefont{E.}~\bibnamefont{Trizac}}, \bibnamefont{and}
  \bibinfo{author}{\bibfnamefont{A.}~\bibnamefont{Prados}},
  \bibinfo{journal}{Phys. Rev. E} \textbf{\bibinfo{volume}{99}},
  \bibinfo{pages}{012140} (\bibinfo{year}{2019}),
  \urlprefix\url{https://link.aps.org/doi/10.1103/PhysRevE.99.012140}.

\bibitem[{\citenamefont{Esposito et~al.}(2010)\citenamefont{Esposito, Kawai,
  Lindenberg, and Van~den Broeck}}]{esposito2010finite}
\bibinfo{author}{\bibfnamefont{M.}~\bibnamefont{Esposito}},
  \bibinfo{author}{\bibfnamefont{R.}~\bibnamefont{Kawai}},
  \bibinfo{author}{\bibfnamefont{K.}~\bibnamefont{Lindenberg}},
  \bibnamefont{and} \bibinfo{author}{\bibfnamefont{C.}~\bibnamefont{Van~den
  Broeck}}, \bibinfo{journal}{EPL (Europhysics Letters)}
  \textbf{\bibinfo{volume}{89}}, \bibinfo{pages}{20003} (\bibinfo{year}{2010}).

\bibitem[{\citenamefont{Diana et~al.}(2013)\citenamefont{Diana, Bagci, and
  Esposito}}]{Esposito_2013}
\bibinfo{author}{\bibfnamefont{G.}~\bibnamefont{Diana}},
  \bibinfo{author}{\bibfnamefont{G.~B.} \bibnamefont{Bagci}}, \bibnamefont{and}
  \bibinfo{author}{\bibfnamefont{M.}~\bibnamefont{Esposito}},
  \bibinfo{journal}{Phys. Rev. E} \textbf{\bibinfo{volume}{87}},
  \bibinfo{pages}{012111} (\bibinfo{year}{2013}),
  \urlprefix\url{https://link.aps.org/doi/10.1103/PhysRevE.87.012111}.

\bibitem[{\citenamefont{Campbell and Deffner}(2017)}]{2017_PRL_Campbell}
\bibinfo{author}{\bibfnamefont{S.}~\bibnamefont{Campbell}} \bibnamefont{and}
  \bibinfo{author}{\bibfnamefont{S.}~\bibnamefont{Deffner}},
  \bibinfo{journal}{Phys. Rev. Lett.} \textbf{\bibinfo{volume}{118}},
  \bibinfo{pages}{100601} (\bibinfo{year}{2017}),
  \urlprefix\url{https://link.aps.org/doi/10.1103/PhysRevLett.118.100601}.

\end{thebibliography}

\end{document}


\preprint{APS/123-QED}

\title{Supplementary Information for ``Engineered Swift Equilibration for Arbitrary Geometries"}

\author{Adam G. Frim}%
\author{Adrianne Zhong}%
\author{Shi-Fan Chen}
\author{Dibuyendu Mandal}
\affiliation{%
 Department of Physics, University of California, Berkeley, Berkeley, CA, 94720
}%
\author{Michael R. DeWeese}
\affiliation{%
 Department of Physics, University of California, Berkeley, Berkeley, CA, 94720
}%
\affiliation{%
Redwood Center For Theoretical Neuroscience, University of California, Berkeley, Berkeley, CA, 94720
}%

\date{\today}
\maketitle

\section{High temperature limit of dipole model}
\label{sec:high_T}
Suppose that we are in the high-temperature regime for the dipole model such that $|pE(t)|\ll k_BT$. In this case we can expand
$\exp{\beta pE(t))} \approx 1 + \beta pE(t)\cos\theta$ and the partition function as $Z(t) = 4\pi + \mathcal{O}(\beta p E(t))^2$. We then have, to the lowest order, the probability density as
\begin{equation}
    \rho(\theta,\phi,t) = \frac{1}{4\pi}(1+\beta pE(t)\cos\theta)
\end{equation}
Taking the partial time derivative of the above gives us an equation of the form $\Delta_{S^2}A= -(4\pi)^{-1}\gamma \beta p \dot{E} P_1(\cos \theta) $, where $P_1(x) = x$
is the first Legendre polynomial. Using the identity $\Delta_{S^2}P_\ell(cos\theta) = -\ell(\ell+1)P_\ell(cos\theta) $, we immediately have the result
\begin{equation}
    A = \frac{1}{8\pi} \gamma \beta p \dot{E} \cos\theta
\end{equation}
Note that if the source term $\gamma \partial_t\rho$ contained a constant a $P_0(\cos \theta) = 1$, Poisson's equation
would be un-invertible; however, as the source term must integrate to zero, such terms are disallowed. 

An interesting note: to first order the requisite force can be applied via a potential that can be practically produced
just as a modulation in the original electric field. In particular we have

\begin{equation}
    \vec{F} = \frac{dA}{\rho_{\text{eq}}} \approx -\frac{1}{2} \gamma \beta p \dot{E} \sin\theta
\end{equation}
This is enforceable by modulating the applied electric field by $\Delta E_{\text{ESE}}(t) = - \frac{1}{2} \gamma \beta \dot{E}_0$ where $E_0$ distinguishes the electric field without the additional ESE contribution. If the original electric field were, for example, a sinusoid, this modulation would manifest as an out-of-phase
contribution to the original field, resulting at first order in a small phase shift of the total signal.

\section{Simulation details}
\label{sec:simulation}
When numerically integrating~(17) and (18) of the main text, the ODE becomes stiff and the numerics break down near the coordinate singularities at $\theta = 0, \pi$. Note that this is purely a feature of our local coordinate chart and not of the physics. Therefore, to simulate the system accurately, we must find ways of dealing with these poles. There are a number of schemes, though the conceptually simplest is simply to define a new coordinate chart and transform to this alternative whenever the system is close to a pole in the original coordinates. In this context, we define a primed set of coordinates $(\theta',\phi')$ which are entirely analogous, i.e. a right-handed spherical coordinate system, though whose zenith is now oriented with the original $x$-axis (the coordinate singularity at $\theta = 0$ now lies on the primed $\hat{y}$-axis). For completeness, we include the transformation in Table~\ref{tab:coord_transform}.

\begin{widetext}
\begin{table}[ht]
    \centering
    \begin{tabular}{|c|c|}
        \hline
         Original Coordinates & Transformed Coordinates   \\
          \hline \hline
         $(x,y,z)$ & $(z',x',y')$\\
         \hline
         $\theta$ & 
         $\text{arctan}\left(\frac{\sqrt{\cos^2 \theta'+\sin^2\theta'\sin^2\phi'}}{\sin\theta'\cos\phi'}\right)$\\
         \hline
         $\text{arctan}\left(\frac{\sqrt{\cos^2\theta+\sin^2\theta\cos^2\phi}}{\sin\theta\sin\phi}\right)$ & $\theta'$\\
         \hline
         $\hat{\theta}$ & $\frac{\cos\theta'\sin\phi'}{sqrt{\cos^2\theta'+\sin^2\theta'\cos^2\theta'}}\hat{\theta}'+\frac{\cos\phi'}{sqrt{\cos^2\theta'+\sin^2\theta'\cos^2\theta'}}\hat{\phi}'$\\
         \hline
    \end{tabular}
    \caption{Coordinate transformation used when $0<\theta<\pi/10$, $9\pi/10<\theta<\pi$ and is therefore close to coordinate singularities in spherical coordinates.}
    \label{tab:coord_transform}
\end{table}
\end{widetext}